\documentclass[a4paper]{JHEP3}
\usepackage{epsfig}
\usepackage{graphicx}
\usepackage{amssymb}
\usepackage{latexsym}

\newcommand{\be}{\begin{equation}}
\newcommand{\ee}{\end{equation}}
\newcommand{\bea}{\begin{eqnarray}}
\newcommand{\eea}{\end{eqnarray}}

\title{On string momentum in effective string theories}
\author{N.D. Hari Dass
\\  DAE Raja Ramanna Fellow, Indian Institute of Science, Bangalore \\
 Email: \email{dass@cts.iisc.ernet.in},\email{ndhari.dass@gmail.com}}

\abstract{Centre of mass momentum for strings is an important ingredient in the calculation of the spectrum of string
theories. It is calculated by the N\"other prescription for two types of conformally invariant effective string theories
of the Polchinski-Strominger type. One is the so called Polyakov-Liouville theory \cite{liouville} and the other an
extension of the original Polchinski-Strominger action analysed upto order $R^{-3}$ where $2\pi R$ is the length 
of the string. In the first case analysis is carried out to order $R^{-2}$ and in the second case to order $R^{-3}$.
In both cases the correction to the free bosonic theory result is shown to be of the {\em improvement} type so that the
correction to the total string momentum, obtained by integrating over the spatial coordinate of the string world sheet, vanishes.
}

\keywords{String Theories, Effective String Theories, String Momentum}

\begin{document}

\section{Introduction}
String-like defects or solitons occur in a wide variety of physical systems. Some well-known examples are vortices
in superfluids, the Nielsen-Olesen vortices of quantum field theories, vortices in Bose-Einstein condensates and
QCD-strings(for reviews see \cite{bali,kutilat05,ournum}). Depending on the particular physical circumstances, these objects can behave quantum-mechanically. The
challenge then is to find consistent ways of describing string-like objects quantum mechanically in \emph{arbitrary}
dimensions. It should be recalled that fundamental string theories are consistent only in the so called critical
dimensions. Polyakov \cite{polya} gave formulations of string theories in \emph{sub-critical} dimensions; 
his ideas play a central role in what follows.
It would be an overkill, and in all likelihood even incorrect, to treat the above mentioned string-like defects
as fundamental strings. A more pragmatic approach would be to treat them in some effective manner much as interactions
of pions are so succesfully described in terms of effective field theories without any pretenses about these field theories
being fundamental at all scales.

Two such approaches to effective string theories exist in the literature. One due to L\"uscher and collaborators \cite{lwearly}, is formulated
entirely in terms of the $D-2$ \emph{transverse} degrees of freedom. It is a case where the \emph{gauge} is fixed completely
without any \emph{residual} invariance left. This approach was further developed in \cite{others}. The recent work of Aharony and Karzbrun has followed this approach in addressing
the issue of spectrum of effective string theories to higher orders. We,along with Drummond, have on the other hand followed
the approach pioneered by Polchinski and Strominger \cite{ps}. In the latter approach, the theories are invariant under \emph{conformal
transformations} and the physical states are obtained by requiring that the generators of conformal transformations
annihilate them. These are too are gauge-fixed theories but with leftover residual invariances characterized by conformal
transformations. It is worth emphasizing that the physical basis of both approaches is that the degrees of freedom are 
transverse.

A major goal in both approaches is the calculation of the spectrum, both of the ground state and of the excited states.
Here we focus only on the Polchinki-Strominger formulation. In this formulation the spectrum calculations need two crucial
ingredients; first of these is the on-shell $T_{--},T_{++}$ and the second is the total centre of mass string momentum.
The calculations proceed by working out the Virasoro generators and an appropriate oscillator algebra(as done in free bosonic 
string theories, for example), and then by imposing physical state conditions. Upon using these physical state conditions and
the string momentum, one can compute the 'masses' of various states through $\sqrt{-P_\mu\cdot P^\mu}$ where $\mu$ are the
target space indices \cite{ps}.

Recently we had given several all-order results claiming that to all orders in $R^{_1}$ the spectrum of all classes of effective
string theories are {\em isospectral} to Nambu-Goto theories or in simpler words there are no corrections to the spectrum of
Nambu-goto theories \cite{liouville,drummond,allaction}. In retrospect, what had been shown unambiguously in these works
was that the on-shell $T_{\pm\pm}$ in all classes of effective string theories was such that the $L_0,{\tilde L}_0$ did not
receive any corrections from their values in Nambu-Goto theories. As the physical state conditions were essentially 
determined by these we had concluded that the spectrum itself would be uncorrected. We had overlooked the fact that the
string momentum may itself receive higher order corrections which would then translate as higher order corrections to the
spectrum. We think that this oversight would only affect the excited states, and that our claim about the spectrum being the
same as in Nambu-Goto case should in fact be true for the ground state. This is because the momentum operator being normal 
ordered should not contribute to the ground state. But it may be desirable to look at this from a variety of angles. Clearly, 
there is a long way to go before these issues are settled. But what is gratifying is that upto the orders treated here, the
assumptions made earlier about the momentum still hold.

In this paper we only investigate the string momentum density operators and their properties. We shall
postpone to a later stage the actual use of these momntum operators in the spectrum calculations.

This paper is organized as follows: in sec 2. we review the simplest case of the free bosonic string theory. In sec 3. we
first evaluate the canonical momentum density for the Polyakov-liouville theory. We truncate the exact expression to order $R^{-2}$ where $2\pi R$ is the length of the closed string(we do not treat open strings here).aWe show that to this order the momentum
density is of the {\em improvement type} with vanishing total centre of mass momentum..In section 4. we carry out the calculations for an extended version of the original
Polchinski-Strominger effective action which is valid to order $R^{-3}$. To this order we show that the momentum density is again of the improvement type.

\section{The free bosonic string theory}
The action for the free bosonic string theory is
\be
\label{freeact}
S_0 = \frac{1}{4\pi a^2}\int~~d\tau^+~d\tau^-~\partial_+X^\mu\cdot\partial_-X^\mu
\ee
here $\tau^\pm=\tau\pm\sigma$ with $\tau,\sigma$ being the temporal and spatial coordinates of the world
sheet, $X^\mu$ are the target space coordinates and $4\pi a^2$ acts as the string tension. We also have
\be
\partial_\pm X^\mu = \frac{1}{2}(\partial_\tau X^\mu \pm \partial_\sigma X^\mu)
\ee
the inverse relations being
\be
\partial_\tau X^\mu = \partial_+ X^\mu + \partial_- X^\mu; \partial_\sigma X^\mu = \partial_+ X^\mu - \partial_- X^\mu
\ee
Let us calculate the string momentum density now. The free action $S_0$ is invariant under the global 
transformation $\delta X^\mu(\tau,\sigma) = b^\mu$ where $b^\mu$ is constant. The way to compute the
corresponding conserved quantity is to let $b^\mu$ depend on $\xi=(\tau,\sigma)$(equivalently $\xi=(\tau_+,\tau_-)$) and compute
\be
\delta S_0 = \int b_\mu~p^\mu(\xi)
\ee
It is straightforward to get
\be
p^\mu_\pm(\xi) = \frac{1}{4\pi a^2}~\partial_\pm X^\mu
\ee
An alternative evaluation is to change coordinates from $\tau_\pm$ to $(\tau,\sigma)$ and rewrite the action as
\be
%\label{freeact}
S_0 = \frac{1}{8\pi a^2}\int~~d\tau~d\sigma~[(\partial_\tau X)^2 - (\partial_\sigma X)^2]
\ee
yielding
\be
p_\tau(\xi)^\mu = \frac{1}{4\pi a^2} \partial_\tau X^\mu \quad\quad p_\sigma(\xi)^\mu = -\frac{1}{4\pi a^2} \partial_\sigma X^\mu
\ee
It is $p_\tau^\mu$ that is identified with canonical momentum density. Clearly the two evaluations agree.
An important property of $p_\pm^\mu$ is its relation with the equations of motion(EOM) $E^\mu$ that 
follows from the action. Quite generally
\be
\label{eomdef}
\delta S = \int E^\mu\cdot \delta X^\mu
\ee
When $\delta X^\mu = b^\mu(\xi)$ we have two ways of evaluating $\delta S$:
\bea
\delta S &=& \int b\cdot E \nonumber\\
         &=&\int\{ p_+\cdot \partial_-b + p_-\cdot\partial_+b\}\nonumber\\
         &=&-\int b\cdot \{\partial_- p_+ + \partial_+ p_-\}
\eea
Since the variations are arbitrary
\be
\label{pcons}
\partial_- p_+^\mu + \partial_+ p_-^\mu = -E^\mu
\ee
When fields satisfy the EOM i.e $E=0$, this becomes a conservation law for $p$. In fact the spatial integral 
$\int d\sigma p_\tau$ is a constant of motion. The EOM for $S_0$ is
\be
E_0^\mu = -\frac{1}{2\pi a^2}\partial_{+-} X^\mu
\ee
It is easy to verify that the $p^\mu$ derived earlier satisfies eqn. (\ref{pcons}).

When $X^\mu$ satisfies the EOM $\partial_{+-} X^\mu = 0$, the {\em classical} solution takes the form
\be
\label{classical}
X^\mu_{cl} = R(e_+^\mu \tau^+ + e_-^\mu \tau^-)
\ee
The free theory $S_0$ is obtained after imposing gauge fixing conditions
\be
\label{gaugefix}
\partial_\pm X\cdot \partial_\pm X =0
\ee
resulting in $e_-\cdot e_- = e_+\cdot e_+=0$. The closed string obeys the periodic boundary condition
$X^\mu(\tau,\sigma +2\pi) = X^\mu(\tau,\sigma)$ for all directions other than the direction along which the
classical string stretches, which we take to be along the $\mu =1$ direction. Along this direction $X^1(\tau,\sigma+2\pi) = X^1(\tau,\sigma)+2\pi R$. This imposes the restriction 
$e_+^\mu-e_-^\mu = \delta_{\mu,1}$. Taken together we get an extra condition $e_+\cdot e_- = -\frac{1}{2}$.

The full solution of the EOM is then
\be
X^\mu = X^\mu_{cl} + Y^\mu
\ee
with $\partial_{+-} Y^\mu =0$. This leads to the mode expansion
\be
\label{mode}
Y^\mu(\tau,\sigma) = q^\mu +\{a\alpha_0^\mu\tau^- + ia\sum_{m\ne 0}\frac{\alpha^\mu_m}{m}e^{-im\tau^-}\}
                           +\{a{\bar\alpha}_0^\mu\tau^+ +ia\sum_{m\ne 0}\frac{{\bar \alpha}^\mu_m}{m}e^{-im\tau^+}\}
\ee
The canonical momentum that results is
\be
\label{momfree}
p_\tau^\mu(\tau,\sigma) = \frac{R}{4\pi a^2}(e_+^\mu + e_-^\mu) + \frac{1}{4\pi a}(\alpha_0^\mu + {\bar \alpha}_0^\mu) + \cdots
\ee
The total string momentum $P^\mu$ becomes
\be
P^\mu = \int d\sigma p_\tau = \frac{R}{2a^2}(e_+^\mu + e_-^\mu) + \frac{1}{2a}(\alpha_0^\mu+{\bar\alpha}_0^\mu)
\ee
The terms denoted by dots in eqn.(\ref{momfree}) integrate to zero. Further, the closed string satisfies
$\alpha_0 = {\bar\alpha}_0$.
\section{The Polyakov-Liouville Theory}
As per the Polchinski-Strominger program \cite{ps}, the free action $S_0$ is to be augmented by additional terms
in the action for effective strings so that the conformal anomaly vanishes in arbitrary dimensions. They added
their Polchinski-Strominger action valid to order $R^{-2}$. We treat this case, which is actually valid to order
$R^{-3}$ \cite{drumorig,ouruniv}, in sec. 4.

The Polyakov-Liouville action is
\be
\label{pol-liou}
S_{(2)} = \frac{\beta}{4\pi}\int d\tau^+ d\tau^- \frac{\partial_+ L\cdot \partial_- L}{L^2}
\ee
where $L = \partial_+ X\cdot \partial_-X$. This action is exactly conformally invariant to all orders in $R^{-1}$.

The variation of this action under arbitrary $\delta X^\mu$ is given by
\be
\label{lpvar}
\delta L_{(2)} = L^{-3}\{L\partial_- L\cdot\partial_+(\delta L) + L\partial_+ L\cdot\partial_-(\delta L) - 2\partial_+ L\cdot\partial_- L\cdot\delta L\}
\ee
with $\delta L = \partial_+ b\cdot\partial_- X + \partial_- b\cdot\partial_+ X$. Let us examine the three terms
in eqn.(\ref{lpvar}) one by one.
\be
\label{lpvar1}
\delta^{(1)} L =\frac{\partial_- L}{L^2}\{\partial_{++}b\cdot\partial_-X+\partial_+b\cdot\partial_{+-}X+
\partial_-b\cdot\partial_{++}X+\partial_{+-}b\cdot\partial_+X\}
\ee 
The terms with second derivatives of $b^\mu$ are to be converted to ones with only first derivatives through
partial integration. Here we come across an ambiguity, namely, the $\partial_{+-}b^\mu$ term can be partially 
integrated in two ways yielding two different expressions. But one should recall that in field theory locally
conserved densities like momentum density, stress tensor etc are {\em ambiguous} upto additions of the so
called improvement terms. These improvement terms are of such a structure that they do not change total momentum
etc. The source of such improvement terms in the present calculation is precisely the ambiguity in the partial
integration of the $\partial_{+-} b$ term. 

The way to handle this is introduce an ambiguity parameter $\alpha_1$ and write $\partial_{+-} b = \alpha_1 \partial_{+-}b + (1-\alpha_1)\partial_{+-}b$ and partially integrate the first wrt to $+$-derivative and partially 
integrate the second wrt the $-$-derivative.

Likewise
\be
\label{lpvar2}
\delta^{(2)} L =\frac{\partial_+ L}{L^2}\{\partial_{--}b\cdot\partial_+X+\partial_-b\cdot\partial_{+-}X+
\partial_+b\cdot\partial_{--}X+\partial_{+-}b\cdot\partial_-X\}
\ee 
and we find similar ambiguity terms. In principle the ambiguous parts of eqn.(\ref{lpvar2}) can be handled with
an independent ambiguity parameter $\alpha_2$ but in the end things must be symmetric wrt exchange $+ \leftrightarrow -$. The third term in eqn.(\ref{lpvar}) has no such ambiguities. We now give the final result for the momentum
densities coming from the $S_{(2)}$ action.
\bea
\label{momlp}
\Delta p_+^\mu &=& \frac{\beta}{2\pi L^3}(\partial_+L\cdot\partial_-L - L\partial_{+-}L)\partial_+X^\mu
               +\frac{\beta\alpha_1}{4\pi}\partial_+(\frac{\partial_+X^\mu\cdot\partial_-L}{L^2}-\frac{\partial_-X^\mu\cdot\partial_+L}{L^2})\nonumber\\
\Delta p_-^\mu &=& \frac{\beta}{2\pi L^3}(\partial_+L\cdot\partial_-L - L\partial_{+-}L)\partial_-X^\mu
               +\frac{\beta\alpha_1}{4\pi}\partial_-(\frac{\partial_-X^\mu\cdot\partial_+L}{L^2}-\frac{\partial_+X^\mu\cdot\partial_-L}{L^2})\nonumber\\
\eea
$\alpha_1$ is the ambiguity parameter referred to before. The extra terms in EOM coming from $S_{(2)}$ are given by
\be
\label{eomlp}
E_{(2)}^\mu = \frac{\beta}{2\pi}[\partial_+\{\partial_-X^\mu(\frac{\partial_{+-}L}{L^2}-\frac{\partial_+L\cdot\partial_-L}{L^3})\}
+\partial_-\{\partial_+X^\mu(\frac{\partial_{+-}L}{L^2}-\frac{\partial_+L\cdot\partial_-L}{L^3})\}]
\ee
It is easy to see that the additional contributions satisfy the conservation equation
\be
\partial_+\Delta p_-^\mu + \partial_- \Delta p_+^\mu = -E_{(2)}^\mu
\ee
As anticipated  the $\alpha_1$-dependent terms in eqn.(\ref{momlp}) are of the improvement type as can be seen
by rewriting them as
\be
\label{improvelp}
\Delta {p^{(\alpha_1})}_\alpha^\mu = \epsilon_{\alpha\beta} \partial^\beta B^\mu \quad\quad B^\mu = \frac{\beta}{4\pi L^2}(\partial_+X^\mu\partial_-L - \partial_-X^\mu\partial_+L)
\ee
and it satisfies the momentum conservation law
\be
\label{improvelpcons}
\partial_+\Delta{p^{(\alpha_1})}_-^\mu + \partial_-\Delta{p^{(\alpha_1})}_+^\mu=0
\ee
without requring $X^\mu$ to satisfy the EOM. This is what is expected of the improvement type terms.

Finally
\be
\label{Lvalue}
L = -\frac{R^2}{2}+R(e_+\cdot\partial_-Y + e_+\cdot\partial_+Y) + \partial_+Y\cdot\partial_-Y
\ee
Consequently only the improvement term in the momentum starts off as $R^{-2}$. Thus to this order the total centre 
of mass momentum of the string vanishes for all states.To this order the EOM reads
\be
\label{eomr2}
-\frac{1}{2\pi a^2}\partial_{+-}Y^\mu + \frac{2\beta}{\pi R^2}[e_-^\mu(e_+\cdot\partial_{++--}Y+e_-\cdot\partial_{+++-}Y)
+e_+^\mu(e_+\cdot\partial_{+---}Y+e_-\cdot\partial_{++--}Y)]=0
\ee
Consequently to this order the solution to the EOM still satisfies $\partial_{+-}Y=0$ and its general form is
\be
Y^\mu(\tau,\sigma) = F^\mu(\tau^+) + G^\mu(\tau^-)
\ee
We record the expressions for the on-shell momentum density to this order(leaving out the classical part)
\bea
\label{momlpr2}
p_+^\mu &=& \frac{1}{4\pi a^2} \partial_+ F^\mu -\frac{\beta\alpha_1}{\pi R^2}e_-^\mu e_-\cdot\partial_{+++}F\nonumber\\
p_-^\mu &=& \frac{1}{4\pi a^2} \partial_- G^\mu -\frac{\beta\alpha_1}{\pi R^2}e_+^\mu e_+\cdot\partial_{---}G
\eea
\section{Polchinski-Strominger Action}
The Polchinski-Strominger action is given by
\be
\label{psact}
S_{PS} = \frac{\beta}{4\pi}\int \frac{(\partial_{++}X\cdot\partial_-X)(\partial_{--}X\cdot\partial_+X)}{L^2}
\ee
%\end{document}
This action is conformally invariant only to order $R^{-3}$ \cite{drumorig,ouruniv}. It differs from $S_{(2)}$ by
terms proportional to $\partial_{+-}X^\mu$ which is just the EOM of $S_0$. Thus the two actions are related by
a small field redefinition. This makes the conformal invariance only approximate and the transformation law is
more complicated looking \cite{ps}.

As before the N\"other prescription gives the modifications, without any ambiguities, to the momentum density as
\bea
\label{momps}
\Delta {p^{(ps)}}^\mu_+ &=& \frac{\beta}{4\pi}[\frac{\partial_{++}X^\mu\partial_{--}X\cdot\partial_+X}{L^2}-\frac{2}{L^3}\partial_+X^\mu\partial
_{++}X\cdot\partial_-X\partial_{--}X\cdot\partial_+X -\partial_-(\frac{\partial_+X^\mu\partial_{++}X\cdot\partial_-X}{L^2})]\nonumber\\
\Delta {p^{(ps)}}^\mu_- &=& \frac{\beta}{4\pi}[\frac{\partial_{--}X^\mu\partial_{++}X\cdot\partial_-X}{L^2}-\frac{2}{L^3}\partial_-X^\mu\partial
_{--}X\cdot\partial_+X\partial_{++}X\cdot\partial_-X -\partial_+(\frac{\partial_-X^\mu\partial_{--}X\cdot\partial_+X}{L^2})]\nonumber\\
\eea
The extra terms in the EOM coming from $S_{PS}$ are
\bea
\label{eomps}
E^\mu_{PS}=\frac{\beta}{4\pi}&[&\partial_{++}(\frac{\partial_-X^\mu\partial_{--}X\cdot\partial_+X}{L^2})
+\partial_+(\frac{2\partial_{++}X\cdot\partial_-X\partial_{--}X\cdot\partial_+X\partial_-X^\mu}{L^3})\nonumber\\
& &-\partial_+(\frac{\partial_{--}X^\mu\partial_{++}X\cdot\partial_-X}{L^2})+ (+\leftrightarrow -)]
\eea
Now we carry out the analysis to order $R^{-3}$ as that is the leading order when momentum gets corrected. At this order the solution
to EOM no longer has the form $Y=F(\tau^+)+G(\tau^-)$; instead it has the general form \cite{liouville}
\be
\label{holosplit}
Y^\mu = F^\mu(\tau^+) + G^\mu(\tau^-) + H^\mu(\tau^+,\tau^-)
\ee
In eqn.(\ref{holosplit}) $H$ is such that it has no purely holomorphic or anti-holomorphic pieces. To order $R^{-3}$ it turns out
\be
\label{hpsr3}
H^\mu = \frac{2\beta a^2}{R^3}[-e_-\cdot\partial_{++}F \partial_-G^\mu-e_+\cdot\partial_{--} \partial_+ F^\mu + e_+^\mu \partial_+\cdot\partial_{--}G + e_-^\mu \partial_{++} F\cdot\partial_-G]
\ee
Because of eqn.(\ref{holosplit}) and eqn.(\ref{hpsr3}) the momentum density to order gets an additional contribution than what one would
have purely from eqn.(\ref{momps}). Gathering all the contributions together one arrives at the total expression for the momentum density
to order $R^{-3}$:
\bea
\label{mompsr3}
{p^{(ps)}}^\mu_+ &=& \frac{\partial_+ F^\mu}{4\pi a^2}
+\frac{\beta}{2\pi R^3}\{
e_-^\mu\partial_{+++}F\cdot\partial_-G-e_+^\mu\partial_{++}F\cdot\partial_{--}G 
- \partial_- G^\mu e_-\cdot\partial_{+++}F + \partial_{++} F^\mu e_+\cdot\partial_{--}G\}\nonumber\\
{p^{(ps)}}^\mu_- &=& \frac{\partial_- G^\mu}{4\pi a^2}
+\frac{\beta}{2\pi R^3}\{
e_+^\mu\partial_{---}G\cdot\partial_+F-e_-^\mu\partial_{--}G\cdot\partial_{++}F 
- \partial_+ F^\mu e_+\cdot\partial_{---}G + \partial_{--} G^\mu e_-\cdot\partial_{++}F\}\nonumber\\
\eea
Miraculously (or the author is too dense to realize whats happening!) the $\beta$-dependent terms can again be written as an
improvement term! Writing these additional terms as ${\bar \Delta}{p^{(ps)}}^\mu$ we have
\bea
{\bar\Delta}{p^{(ps)}}^\mu_+ = \frac{\beta}{2\pi R^3}\partial_+ C^\mu\nonumber\\
{\bar\Delta}{p^{(ps)}}^\mu_- = -\frac{\beta}{2\pi R^3}\partial_- C^\mu\nonumber\\
\eea
where
\be
C^\mu = e_-^\mu \partial_{++}F\cdot\partial_-G -e_+^\mu \partial_+F\cdot\partial{--}G + E_+\cdot\partial_{--}G \partial_+ F^\mu
-e_-\cdot\partial_{++}F \partial_{--} G^\mu
\ee
This means that to order $R^{-3}$ the total centre of mass momentum for all states does not get any corrections over the free theory value!
\section{Discussions and Conclusions}
We have computed the string momentum density for the Liouville-Polyakov effective string theory \cite{liouville}
as well as for the Polchinski-Strominger action \cite{drumorig,ouruniv,liouville}. By the N\"other procedure we
have obtained expressions which are valid to all orders in $R^{-1}$ in the case of the former, and to order $R^{-3}$
in the case of the latter. In the case of Polyakov-Liouville case we found part of the answer to be of the improvement
type to all orders. The analysis was then restricted to $R^{-2}$ order whence only the improvement term matters and the
correction to the total string momentum, obtained by integrating the canonical momentum density, coming from the 
Polyakov-Liouville action, over the spatial volume of the worldsheet,
vanishes for all states. In the Polchinski-Strominger case too, the correction to the canonical momentum density is
shown to be of the improvement type once again resulting in vanishing total string momentum.

These results are encouraging and assume significance in the light of the fact that our earlier all-order proofs had
overlooked this important issue of possible corrections to total string momentum. It remains a challenge to extend
the results of this current work to all orders for all possible classes of effective string theories as constructed by our
Covariant Calculus \cite{covariant}. It should however be stressed that for the ground state the normal ordering of the
canonical  momentum density ensures that our earlier considerations should remain valid.

\acknowledgments

The author would like to express his gratitude to the Department of Atomic Energy for the award of a
Raja Ramanna Fellowship which made this work possible, and to CHEP, IISc for its invitation to use
this Fellowship there. He also expresses his sincere thanks to Yashas Bharadwaj who participated in the
early phases of this work and for his many insightful inputs. His Junior Research Fellowship was also supported
under the DAE Raja Ramanna Fellowship scheme.

\end{document}